\documentclass[useAMS,usenatbib,usegraphicx]{mn2e}

\usepackage{times}

\newcommand{\gps}{\ensuremath{g_{\rm P1}}}
\newcommand{\rps}{\ensuremath{r_{\rm P1}}}
\newcommand{\ips}{\ensuremath{i_{\rm P1}}}
\newcommand{\zps}{\ensuremath{z_{\rm P1}}}
\newcommand{\yps}{\ensuremath{y_{\rm P1}}}
\newcommand{\grizy}{\gps\rps\ips\zps\yps}
\newcommand{\PS}{\protect \hbox {Pan-STARRS1}}

\title[Fiducials in the \PS\ Photometric System]
      {Galactic Globular and Open Cluster Fiducial Sequences in the
       \PS\ Photometric System}
\author[E.~J.\ Bernard et al.]{%
Edouard J. Bernard,$^{1}$\thanks{E-mail: ejb@roe.ac.uk}
Annette M. N. Ferguson,$^{1}$
Edward F. Schlafly,$^{2}$
\newauthor
Imants Platais,$^{3}$
Eric F. Bell,$^{4}$
Nicolas F. Martin,$^{5,2}$
Hans-Walter Rix,$^{2}$
\newauthor
Colin T. Slater,$^{4}$
William S. Burgett,$^{6}$
Kenneth C. Chambers,$^{6}$
Peter W. Draper,$^{7}$
\newauthor
Klaus W. Hodapp,$^{6}$
Nicholas Kaiser,$^{6}$
Rolf-Peter Kudritzki,$^{6}$
Eugene A. Magnier,$^{6}$
\newauthor
Nigel Metcalfe,$^{7}$
John L. Tonry,$^{6}$
Richard J. Wainscoat,$^{6}$
Christopher Waters$^{6}$\\
$^{1}$SUPA, Institute for Astronomy, University of Edinburgh, Royal
   Observatory, Blackford Hill, Edinburgh EH9 3HJ, UK \\
$^{2}$Max-Planck-Institut f\"ur Astronomie, K\"onigstuhl 17, D-69117
   Heidelberg, Germany \\
$^{3}$Department of Physics and Astronomy, Johns Hopkins University,
   3400 North Charles Street, Baltimore, MD 21218, USA \\
$^{4}$Department of Astronomy, University of Michigan, 500 Church St.,
   Ann Arbor, MI 48109, USA \\
$^{5}$Observatoire Astronomique de Strasbourg, Universit\'e de Strasbourg,
   CNRS, UMR 7550, 11 rue de l'Universit\'e, F-67000 Strasbourg, France \\
$^{6}$Institute for Astronomy, University of Hawaii, 2680 Woodlawn Drive,
   Honolulu HI 96822, USA \\
$^{7}$Department of Physics, Durham University, South Road, Durham DH1 3LE, UK
}

\begin{document}

\date{Accepted --. Received --; in original form --}

\pagerange{\pageref{firstpage}--\pageref{lastpage}} \pubyear{2013}

\maketitle

\label{firstpage}

\begin{abstract}

 We present the fiducial sequences of a sample of Galactic star clusters
 in the five bands of the \PS\ (PS1) photometric system (\gps, \rps, \ips,
 \zps, and \yps). These empirical sequences -- which include the red giant
 and sub-giant branches, the main sequence, and the horizontal branch --
 were defined from deep colour-magnitude diagrams reaching below the oldest
 main-sequence turn-offs of 13 globular and 3 old open clusters covering a
 wide range of metallicities ($-2.4\la\rm{[Fe/H]}\la+0.4$). We find
 excellent agreement for the nine clusters in common with previous studies
 in similar photometric systems when transformed to the PS1 system.
 Because the photometric and spectroscopic properties of these stellar
 populations are accurately known, the fiducials provide a solid basis for
 the interpretation of observations in the PS1 system, as well as valuable
 constraints to improve the empirical colour--$T_{\rm {eff}}$ relations.

\end{abstract}


\defcitealias{har96}{H10}

\section{Introduction}

 Since its introduction by the {\it Sloan Digital Sky Survey}
 \citep[SDSS][]{yor00}, the {\it ugriz} photometric system has been widely
 adopted for optical observations, and adapted to the various instruments
 and needs of the surveys. In particular, the recent progress in
 deep-depletion CCD technology enhancing the quantum efficiency at the
 reddest wavelengths
 allows the addition of a $y$ filter, which benefits greatly from being more
 immune to interstellar reddening -- and therefore the possibility to peer
 deeper into the Milky Way disc -- as well as higher temperature
 sensitivity for brown dwarfs studies and better constrained photometric
 redshifts. Thanks to the high throughput of the SDSS filters, it has
 become the reference system for many current and upcoming large sky
 surveys (e.g.\ \PS\ (PS1): \citealt{kai02}, Large Synoptic Survey
 Telescope: \citealt{tys02}, Dark Energy Survey: \citealt{des05}, Subaru
 Hyper Suprime-Cam Project: \citealt{tak10}, VST ALTAS: \citealt{sha13}).

 However, proper interpretation of the observed stellar populations
 depends on understanding the relation between the measured properties
 (e.g.\ colour, magnitude) and the physical properties (e.g.\ temperature,
 metallicity). One possibility is to compare directly the observations to
 stellar systems for which the properties are accurately known, such as
 globular and open clusters (GCs and OCs, respectively). It is
 also possible to use the observations of these clusters as robust
 empirical constraints to improve theoretical models. For example,
 previous studies have highlighted the difficulty of fitting the
 colour-magnitude diagrams (CMDs) of Galactic GCs using the current
 stellar evolution libraries \citep[e.g.][]{van03,dot07,an08,bra10}, and the
 need for semi-empirical colour-temperature relations obtained from
 accurate star cluster fiducials \citep[e.g.][see also \citealt{sal02}]{van03}.

\begin{figure}
\includegraphics[width=8.5cm]{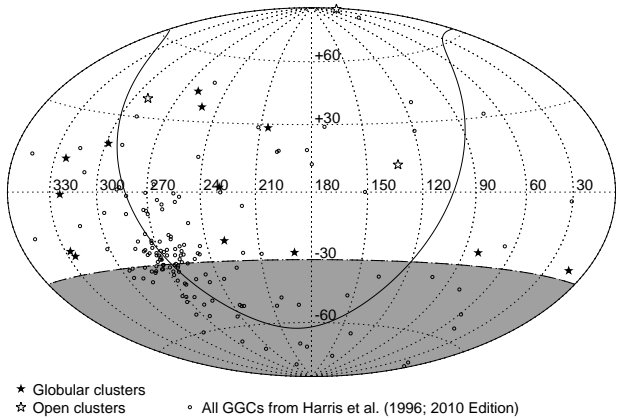}
\caption{Distribution in equatorial coordinates of the globular
 clusters analysed in this work (filled stars), selected from the
 \citetalias{har96} catalogue of Galactic GCs (open circles). The three open
 clusters in our sample are shown as open stars.
 The solid line traces the Galactic disc, while the grayed area represents
 the fraction of the sky not observable from the PS1 3$\pi$ telescope.}
\label{fig:map}
\end{figure}

 We have taken advantage of the recently completed PS1 3$\pi$ Survey
 (see below) to define a library of empirical fiducial sequences in the
 five bands of the PS1 photometric system (\gps, \rps, \ips, \zps,
 and \yps) based on deep observations of a sample of Galactic star
 clusters. Our goal is to use these fiducials for comparison with the
 stellar populations of the Milky Way halo, satellites, and globular
 clusters, hence the focus on old clusters.
 We briefly introduce PS1 in Section~\ref{ps1}, and describe
 the photometry in Section~\ref{phot}. The cluster fiducials are defined
 and compared to literature values in Section~\ref{fid}. A summary is
 given in Section~\ref{summ}.

\section{The \PS\ 3$\pi$ Survey}\label{ps1}

Pan-STARRS1 \citep[K.\,C.\ Chambers et al., in preparation]{kai10} is a
1.8~m optical telescope installed on the peak
of Haleakala (Hawaii) and designed for dedicated survey observations.
It is equipped with a 1.4-Gigapixel imager \citep{ona08,ton09} covering a 7
square degree field-of-view ($\sim3.3\degr$ diameter).
The 3$\pi$ Survey, which forms the basis of the work presented here, covers
3/4 of the sky in five optical to near-infrared bands \citep[\grizy;][]{ton12}.

With an exposure time ranging from 30 to 45 seconds, individual exposures
have median 5$\sigma$ limiting AB magnitudes of 21.9, 21.8, 21.5, 20.7, and
19.7 for the \grizy bands, respectively \citep{mor12}; saturation occurs
at $\sim$13.5 for \gps, \rps, \ips, $\sim$13.0 for \zps, and $\sim$12.0 for
\yps \citep{mag13}. The whole sky visible from Hawaii is being observed four
times per band and per year, over a period of about four years, which
will lead to an increased depth of $\sim$1.2~mag on the final stacked images
\citep{met13}. The median seeing is a function of wavelength, ranging
from 1.0$\arcsec$ in \yps\ to 1.3$\arcsec$ in \gps \citep{met13}.
The individual frames are automatically processed with the Image Processing
Pipeline \citep{mag06} to produce a photometrically and astrometrically
calibrated catalogue. In particular, the images are resampled to a uniform
pixel size (0.25$\arcsec$) and aligned to the equatorial axes on regular areas
on the sky (called {\it skycells}). These skycells are roughly 6250 pixels
across ($\sim$26$\arcmin$), and the resulting images are called {\it warps}.

\begin{figure}
\includegraphics[width=8.5cm]{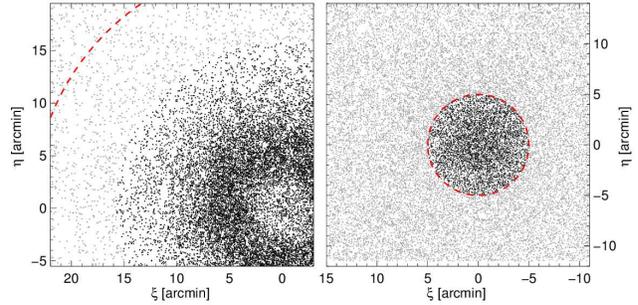}
\caption{Spatial distribution, in standard coordinates, of the stars within
 the analysed skycell for
 globular cluster NGC\,5904 (left) and open cluster NGC\,6791 (right).
 Black and gray points represent the stars used to determine the cluster
 fiducial and the rejected stars, respectively. The dashed circle in
 each panel shows the characteristic radius $r$ (see Table~\ref{tab:props}).}
\label{fig:dist}
\end{figure}

\section{Cluster Photometry}\label{phot}

Using the 2010 Edition of the \citet{har96} catalogue of Galactic GCs
\citep[hereafter H10]{har10}, we count 96 known GCs located in the footprint
of PS1 (i.e.\ $\delta>-30$; see Figure~\ref{fig:map}). Because the objective
of our work is to obtain well-defined cluster fiducials over a wide range
of magnitudes, including the main-sequence turn-off (MSTO), we limited our
analysis to nearby, well-populated, and low foreground reddening clusters.
All the clusters from \citetalias{har96} satisfying the following
constraints were selected:
$(m-M)<16$, ${\rm E}(B-V)\leq0.1$, and ${\rm M}_V<-6$. This reduced the list
to eleven clusters. To extend the range of metallicities covered by the
sample, we added NGC\,6838 (M\,71) and Pal\,12, as well as the old open
clusters NGC\,188, NGC\,6791, NGC\,2682 (M\,67). The properties of the whole
sample are summarised in Table~\ref{tab:props}.

Because the Survey data reduction is still ongoing, neither the reduced images
nor the photometric catalogues are final. At the beginning of this project the
quality of the pipeline photometry in crowded regions was not optimal. It was
thus decided to perform the stellar photometry on the PS1 images with the
standard {\sc daophot/allstar/allframe} suite of programs \citep{ste94}, which
was specifically developed for crowded field photometry.

\begin{figure*}
\includegraphics[width=14cm]{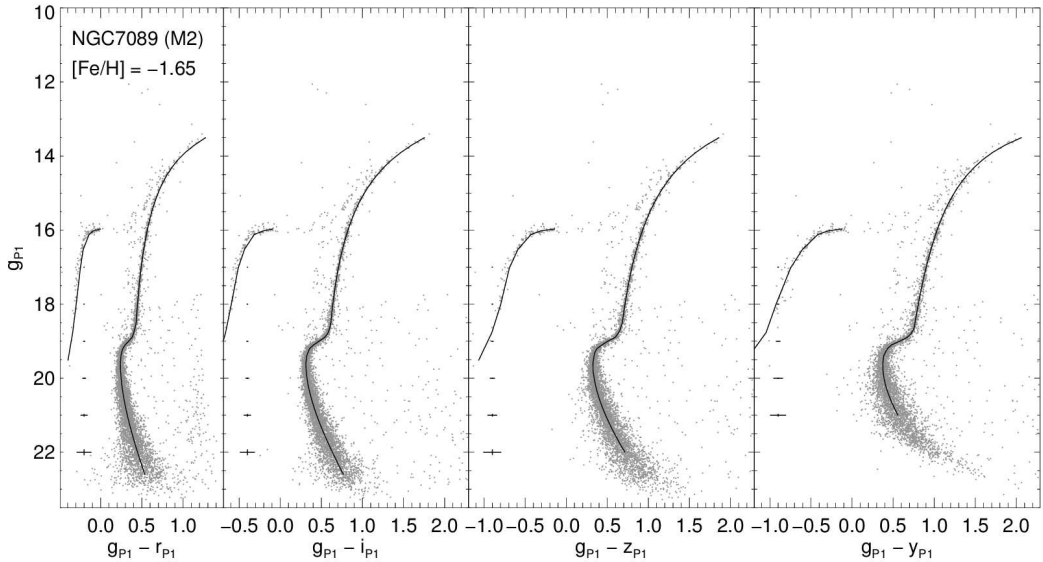}
\caption{CMDs and corresponding fiducials for globular cluster NGC\,7089.}
\label{fig:fid}
\end{figure*}

\begin{table*}
\centering
 \begin{minipage}{135mm}
\caption{Properties of the globular and open clusters in our sample, from \citetalias{har96} (GCs) and \citet[OCs]{dia02}.}
\label{tab:props}
\begin{tabular}{ @{}cccccccc}
  \hline
   Name     &Alt. Name&   RA (J2000)  &   DEC (J2000)   &  [Fe/H]   &E(B$-$V)&(m$-$M)& $r$\footnote{the characteristic radius $r$ represents the tidal radius for the GCs, and the apparent radius for the OCs.} (arcmin)  \\
  \hline
 NGC\,288   &   ...   &  00 52 45.24  &  $-$26 34 57.4  &  $-$1.32  &  0.03  &  14.84    & 13.2 \\
 NGC\,1904  &  M\,79  &  05 24 11.09  &  $-$24 31 29.0  &  $-$1.60  &  0.01  &  15.59    &\ 8.0 \\
 NGC\,4590  &  M\,68  &  12 39 27.98  &  $-$26 44 38.6  &  $-$2.23  &  0.05  &  15.21    & 14.9 \\
 NGC\,5272  &  M\,3   &  13 42 11.62  &  $+$28 22 38.2  &  $-$1.50  &  0.01  &  15.07    & 28.7 \\
 NGC\,5897  &   ...   &  15 17 24.50  &  $-$21 00 37.0  &  $-$1.90  &  0.09  &  15.76    & 10.1 \\
 NGC\,5904  &  M\,5   &  15 18 33.22  &  $+$02 04 51.7  &  $-$1.29  &  0.03  &  14.46    & 23.6 \\
 NGC\,6205  &  M\,13  &  16 41 41.24  &  $+$36 27 35.5  &  $-$1.53  &  0.02  &  14.33    & 21.0 \\
 NGC\,6341  &  M\,92  &  17 17 07.39  &  $+$43 08 09.4  &  $-$2.31  &  0.02  &  14.65    & 12.4 \\
 NGC\,6838  &  M\,71  &  19 53 46.49  &  $+$18 46 45.1  &  $-$0.78  &  0.25  &  13.80    &\ 8.9 \\
 NGC\,7078  &  M\,15  &  21 29 58.33  &  $+$12 10 01.2  &  $-$2.37  &  0.10  &  15.39    & 27.3 \\
 NGC\,7089  &  M\,2   &  21 33 27.02  &  $-$00 49 23.7  &  $-$1.65  &  0.06  &  15.50    & 12.4 \\
 NGC\,7099  &  M\,30  &  21 40 22.12  &  $-$23 10 47.5  &  $-$2.27  &  0.03  &  14.64    & 19.0 \\
 Pal\,12    &   ...   &  21 46 38.84  &  $-$21 15 09.4  &  $-$0.85  &  0.02  &  16.46    & 19.1 \\
  \hline
 NGC\,188   &   ...   &  00 47 28     &  $+$85 15 18    &  $-$0.03  &  0.08  &  11.56    &  8.5 \\
 NGC\,2682  &  M\,67  &  08 51 18     &  $+$11 48 00    &  $+$0.03  &  0.04  & \ 9.49    & 12.5 \\
 NGC\,6791  &   ...   &  19 20 53     &  $+$37 46 18    &  $+$0.42  &  0.16  &  13.51    &  5.0 \\
  \hline \vspace{-10mm}
\end{tabular}
\end{minipage}
\end{table*}


For each cluster (except NGC\,188, see below) a single skycell was analysed;
a region of $\sim$26' on a
side is sufficient to sample a significant fraction of the area within the
tidal radius of any cluster, regardless of their precise location
within the skycell. Figure~\ref{fig:dist} shows the spatial distribution of
stellar objects within the studied skycell for an extended (NGC\,5904) and
a more compact (NGC\,6791) cluster.
All the warps in all five bands of a given skycell were retrieved from the
PS1 Science Interface\footnote{http://web01.psps.ifa.hawaii.edu}. Between
2 and 25 images per band were available for each cluster, with a median of
12 images per band.

We performed a first source detection at the 10-$\sigma$ level on the
individual warps, which was used as input for aperture photometry. From
this catalogue, 200 bright, non-saturated stars per warp were selected as
potential PSF stars; an automatic rejection based on the shape
parameters was used to clean the lists. Modelling of the empirical PSF was
done iteratively with {\sc daophot}: the clean lists were used to remove all
the stars from the warps except PSF stars, so that accurate PSFs could be
created from non-crowded stars. At each iteration, the PSF was modeled
more accurately and the neighbouring stars thus removed better. Every few
iterations, the degree of PSF variability across the image was also
increased, from constant to linear, then quadratically variable.

We then created a median, master image by stacking all the warps from all
the bands using the stand-alone program {\sc montage2}. This master image,
free of chip gaps and much deeper than any individual frame, was used to
create the input star list for {\sc allframe} by performing a second source
detection. The output of {\sc allframe} consists of a catalogue of PSF
photometry for each image. A robust mean magnitude was obtained for each
star by combining these catalogues with {\sc daomaster}.
The final photometry was calibrated to the PS1 system by matching our
resulting catalogue to that produced and calibrated by the pipeline
\citep[see][]{sch12} for the given skycell. Using several hundreds of
stars in common in the uncrowded areas of the skycells, the residuals
have a standard deviation $\la$0.02~mag and show no trend with either
magnitude or colour. The contribution of these transformations to the
uncertainty of the individual magnitudes is therefore negligible
($\sim$0.001).

\begin{figure}
\includegraphics[width=8.cm]{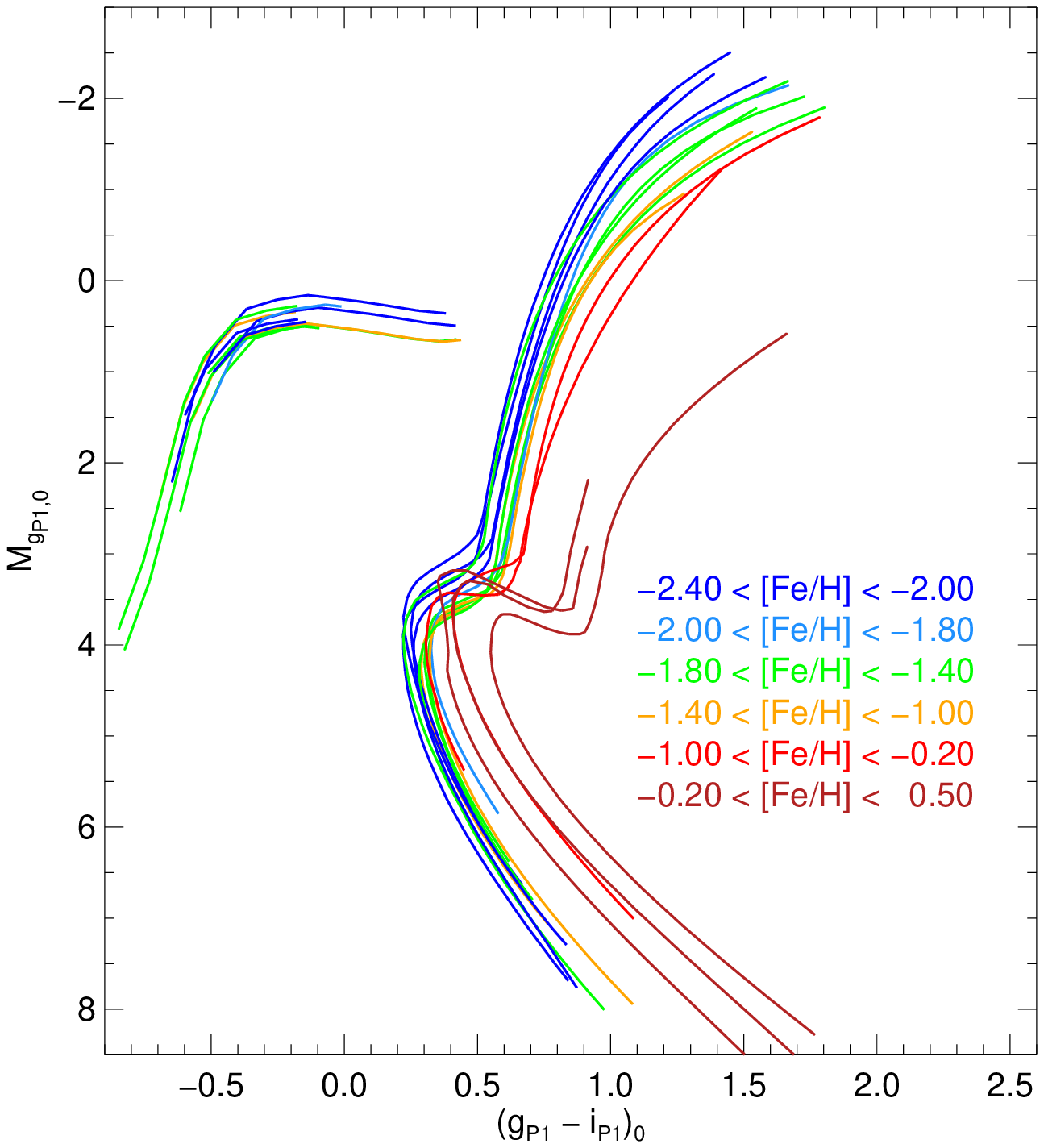}
\includegraphics[width=8.cm]{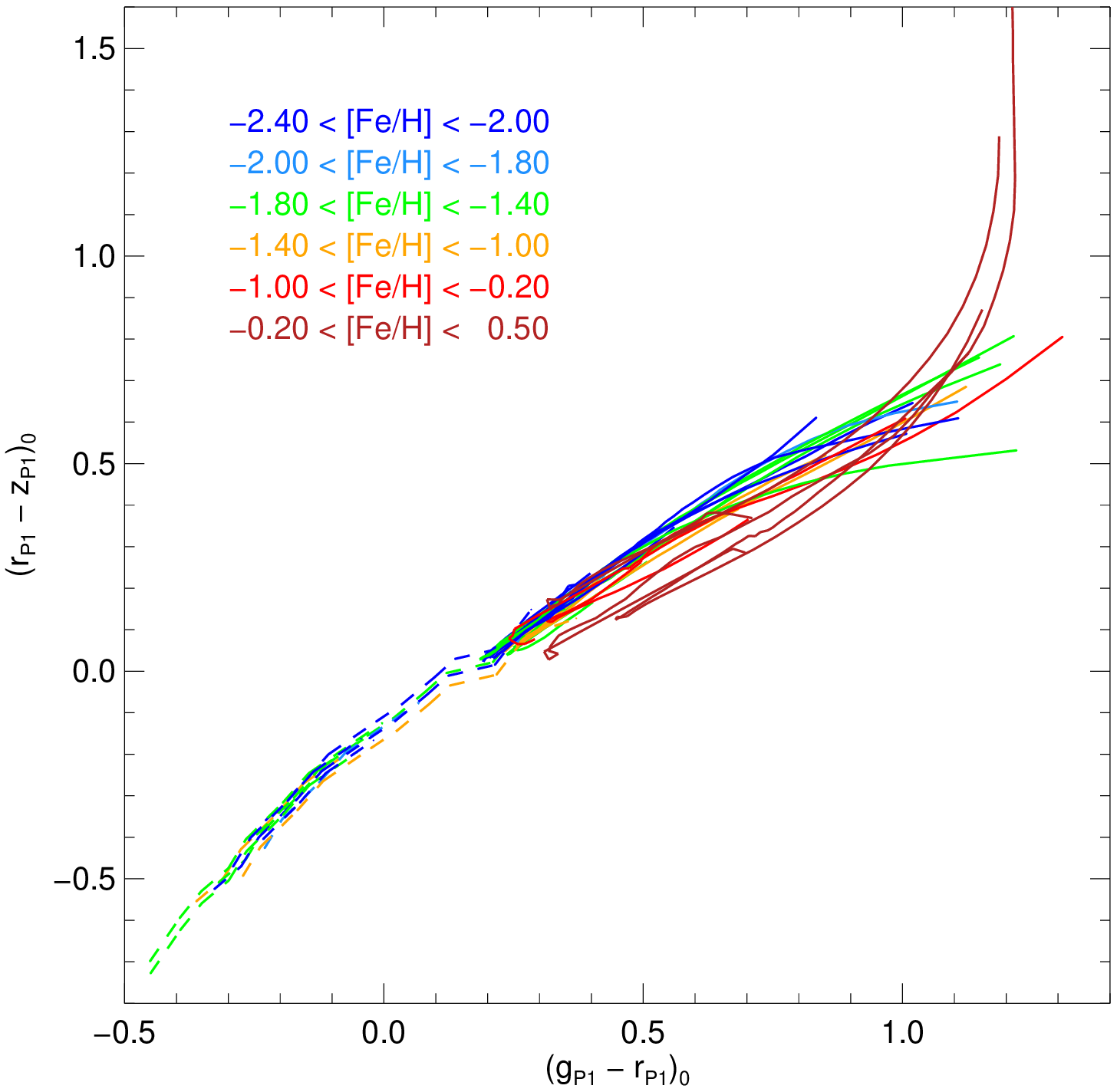}
\caption{Comparison of the fiducials obtained for all the clusters,
 colour-coded by their metallicity as described in the insets, in a
 CMD (top) and colour-colour diagram (bottom). In the latter, the HBs
 are shown as dashed lines for clarity.}
\label{fig:fid_vs_FeH}
\end{figure}

The photometry of NGC\,188 was performed in a slightly different manner
from the other clusters. It is located in a part of the sky for which the
data was not fully processed by the pipeline yet, leading to two minor
differences: (i) the warps were not available, so the photometry was carried
out on the un-resampled, un-rotated images of the cluster; and (ii) the
photometric calibration is preliminary and may not be as accurate as for
the other clusters.

The final step consisted of cleaning the photometric catalogues of
non-stellar objects and blended stars in order to obtain a well-defined
stellar locus. This was possible thanks to the photometric quality
parameters returned by {\sc allframe} ($\chi$ and {\it sharp}),
as well as the separation index {\it sep} \citep[see][]{ste03} to reduce
the effects of photometric degradation due to crowding.
Specifically, we used $|{\rm \it sharp}|\leq1.5$, {\it sep}~$\geq3.5$,
and a magnitude dependent $\chi$. To limit the number of foreground
contaminants, we only included objects within $\sim$1/2--1 characteristic
radius $r$ (see Table~\ref{tab:props}), depending on the density of
surrounding field stars. The CMDs of the open clusters NGC\,188 and
NGC\,6791 were further cleaned by removing the foreground Milky Way
stars based on their proper motions \citep{pla03,pla11}.

\section{Cluster Fiducials}\label{fid}

\subsection{Defining the Cluster Fiducials}

To determine the cluster fiducials, we found that the ($\gps-X, \gps$)
plane, where X represents one of the \rps, \ips, \zps, or \yps\ filters,
produced the best results since the MSTO is better defined by including
bluer bands. Sample CMDs corresponding to these band combinations are shown
in Figure~\ref{fig:fid}.

For the red giant branch (RGB), sub-giant branch (SGB), and main sequence
(MS), the fiducials were defined in a manner similar to that described in
\citet{mar09}. First, the sigma-clipped median colour is calculated for
magnitude bins that have a size which varies as a function of the number
of stars and the photometric errors, from 0.2 to 0.8~mag. These values
define the preliminary ridge line. This is then refined between the MSTO
and the base of the RGB by computing, at each ridge point, the distribution
of stars in a stripe perpendicular to the ridge line. To obtain smoother
fiducials, we separately fit the RGB and the MS below the turn-off
with a function of the form:
\[
y = a + b x + c/(x - d)
\]
where x and y represent the magnitude and the colour, respectively
\citep[see][]{sav00}\footnote{Note that \citet{sav00} used the reverse
notation, with $x$ and $y$ representing the colour and the magnitude,
respectively}. Finally, for the few clusters where the cluster locus is
double valued in color for a given magnitude (e.g.\ the SGB of the open
clusters), the fiducial was refined by hand.

\begin{figure*}
\includegraphics[width=17cm]{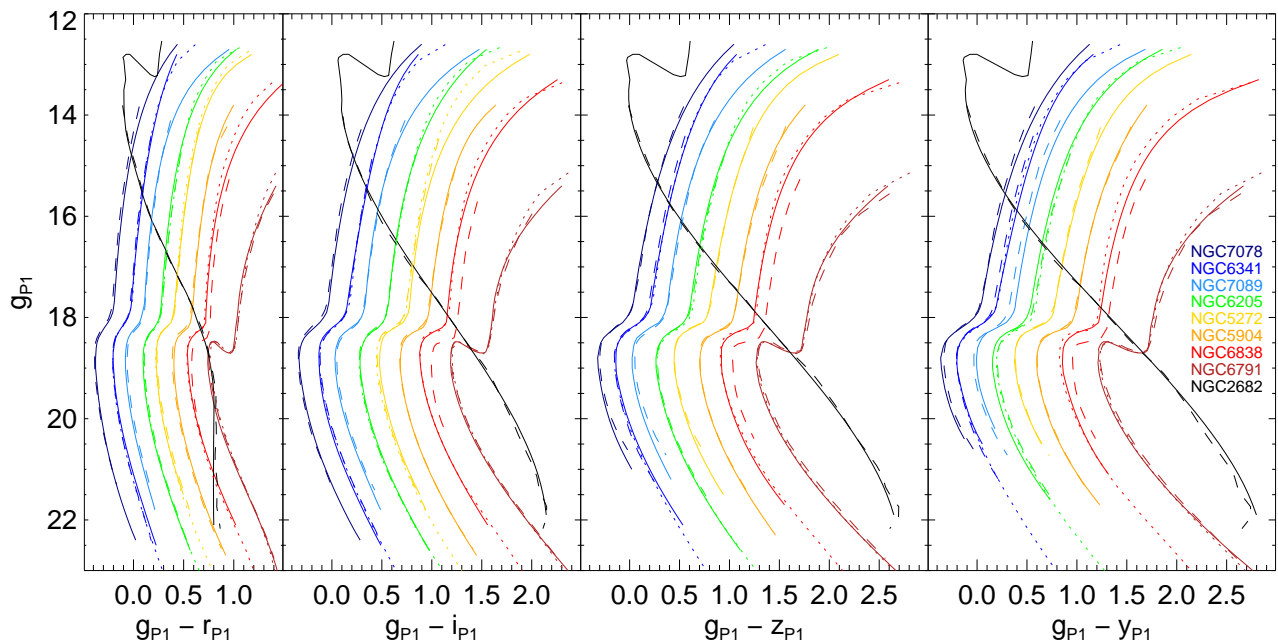}
\caption{Comparison between the fiducials obtained in this work (solid lines)
 and those of the literature after transformation to the PS1 system:
 Clem et al.\ (dotted lines) and An et al.\ (dashed lines). The fiducials of
 individual clusters have been arbitrarily offset in colour and magnitude for
 clarity.}
\label{fig:comp_fid}
\end{figure*}

Given the typically small number of stars on the horizontal-branch (HB)
of individual clusters, we first created a robust HB fiducial by combining
the HB stars of all the clusters together after correcting for the
difference in distance and reddening. For each cluster, this fiducial was
then simply offset in colour and magnitude, then trimmed to match the
observed HB. Figure~\ref{fig:fid} shows the fiducials obtained in the
various band combinations for NGC\,7089, overplotted on the cleaned CMDs.
The fiducials for the other clusters are shown in the Appendix, and
are provided in tabular form available as Supporting Information with the
online version of the paper.

In Figure~\ref{fig:fid_vs_FeH} we present the fiducials of all the clusters
colour-coded by their metallicity: the top and bottom panels show a
CMD and a colour-colour diagram, respectively. The fiducials were converted
to absolute magnitude and intrinsic colour assuming the distances and
reddening from \citetalias{har96} and \citet{dia02}, and the extinction in
each band from \citet{sch11}. Note that the bright end of several fiducials
do not include the tip of the RGB as stars brighter than $\gps\sim 13.5$ are
saturated in the PS1 images. The CMDs clearly show the expected trend of
redder colour and fainter sub-giant branch with increasing metallicity. A
mild metallicity trend is also visible in the bottom panel; since this
diagram is distance-independent, it may prove useful for constraining the
metallicity of field stars.

\subsection{Comparison with Literature Fiducials}

Given that the filter set has been designed specifically for PS1, there
are no fiducials in the literature that are directly comparable to ours.
However, fiducials have been obtained for a number of clusters in the
SDSS $ugriz$ and CFHT/MegaCam $u'g'r'i'z'$ systems. The similarity of
the PS1 filters to these allows the determination of robust
transformations between the systems \citep[see][]{ton12}.

The first extensive set of fiducials was obtained by \citet{cle08}. They
secured very deep observations of four globular clusters and one open
cluster with the CFHT. The combination of long and short exposure times
allowed them to obtain high quality fiducials ranging from the tip of the
RGB to roughly 4 magnitudes below the MSTO. All the clusters they analysed
are included in our sample.
The work of \citet{an08} was carried out with the SDSS images. Since the
standard SDSS photometric pipeline could not handle the high stellar
density in and around Milky Way clusters, \citet{an08} re-analysed the
images around the 17 globular clusters found in the SDSS footprint, as
well as three open clusters, with {\sc daophot/allframe}. Seven globular
and two open clusters are in common with ours.

We first converted the \citet{cle08} fiducials to the SDSS photometric
system using the transformations from \citet{tuc06}; both sets of
fiducials were then transformed to the PS1 system using Equation~6 in
\citet{ton12}. Note that the SDSS and CFHT filter sets do not have a $y$
filter, so this is extrapolated from the $z$-band \citep[see][]{ton12}.
Finally, we corrected the SDSS and CFHT $r$-band magnitudes for a
systematic offset, in the sense that they appear too faint by 0.03~mag
\citep[see also \citealt{fuk11}]{ton12}.

The comparison is shown in Figure~\ref{fig:comp_fid}. The fiducials defined
in this work are shown as solid lines, while the ones from \citet{cle08} and
\citet{an08} are shown as dashed and dotted lines, respectively. For
magnitudes fainter than $\gps \sim 14$, we find excellent agreement
between the various sets (i.e.\ better than 0.02~mag in \gps, \rps, \ips,
and \zps, and better than 0.06 ~mag in the \yps-band) indicating that the
photometric calibration and transformations between systems are accurate.
The only exception is for NGC\,6838 (M\,71). \citet{an08} already noted a
$\sim$0.1~mag offset to the red compared to the fiducials of \citet{cle08},
citing the difficulty to obtain a reliable calibration in a field with
high stellar crowding which prevented the SDSS pipeline to measure a
sufficient number of stars. Our fiducials for this cluster agree with the
ones of \citet{cle08} and therefore confirm the problematic calibration
of the NGC\,6838 photometry of \citet{an08}.

\section{Conclusions}\label{summ}

We have obtained deep, homogeneous photometry of a sample of globular
and open clusters covering a wide range of metallicities from images of the
PS1 3$\pi$ Survey. We used this photometry to derive fiducial sequences in
the five bands of the PS1 photometric system. The comparison with literature
fiducials in similar photometric systems shows very good agreement and
therefore that the photometric calibration and transformations between
systems are accurate.

These empirical isochrones, for which the photometric and spectroscopic
properties are accurately known, can be used to characterize the properties
of resolved stellar systems found in the footprint of PS1 and other surveys
observing with the same filters, as well as to improve the empirical
colour--$T_{\rm {eff}}$ relations predicted from model atmospheres.

\section*{Acknowledgments}

We are grateful to the anonymous referee for a constructive report. This
research was supported by a consolidated grant from the Science Technology
and Facilities Council. E.F.S. and N.F.M. acknowledge support from the DFG's
grant SFB881 (A3) ``The Milky Way System". N.F.M. gratefully acknowledges the
CNRS for support through PICS project PICS06183.

The PS1 Surveys have been made possible through contributions of
the Institute for Astronomy, the University of Hawaii, the Pan-STARRS Project
Office, the Max-Planck Society and its participating institutes, the Max Planck
Institute for Astronomy, Heidelberg and the Max Planck Institute for
Extraterrestrial Physics, Garching, The Johns Hopkins University, Durham
University, the University of Edinburgh, Queen's University Belfast, the
Harvard-Smithsonian Center for Astrophysics, the Las Cumbres Observatory Global
Telescope Network Incorporated, the National Central University of Taiwan, the
Space Telescope Science Institute, the National Aeronautics and Space
Administration under Grant No.\ NNX08AR22G issued through the Planetary Science
Division of the NASA Science Mission Directorate,  the National Science
Foundation under Grant No.\ AST-1238877, and the University of Maryland.

\appendix

\section{Fiducials}

\begin{figure*}
\includegraphics[width=12.5cm]{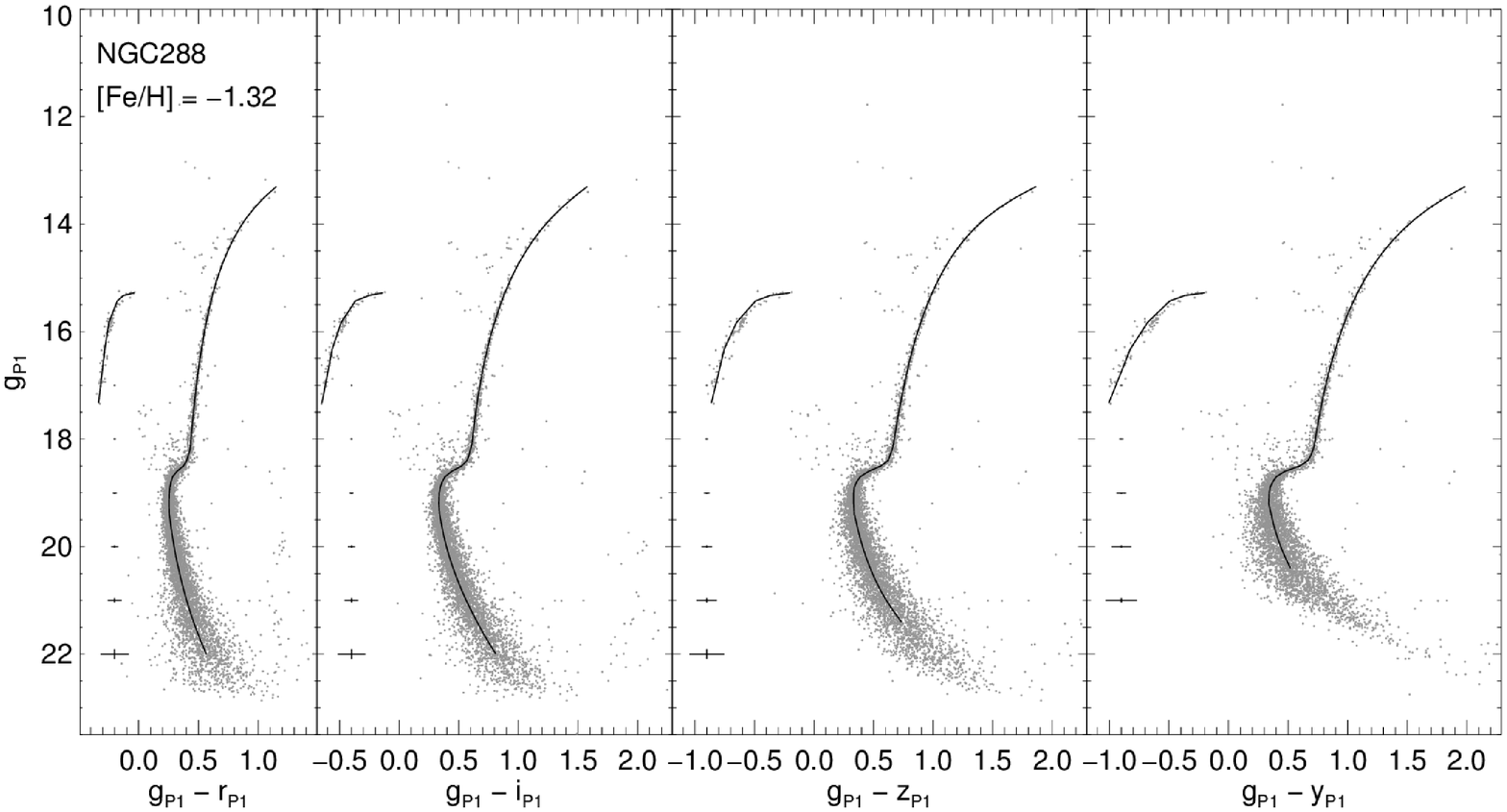}
\caption{CMDs and corresponding fiducials for globular cluster NGC\,288.}
\label{fig:f288}
\end{figure*}

\begin{figure*}
\includegraphics[width=12.5cm]{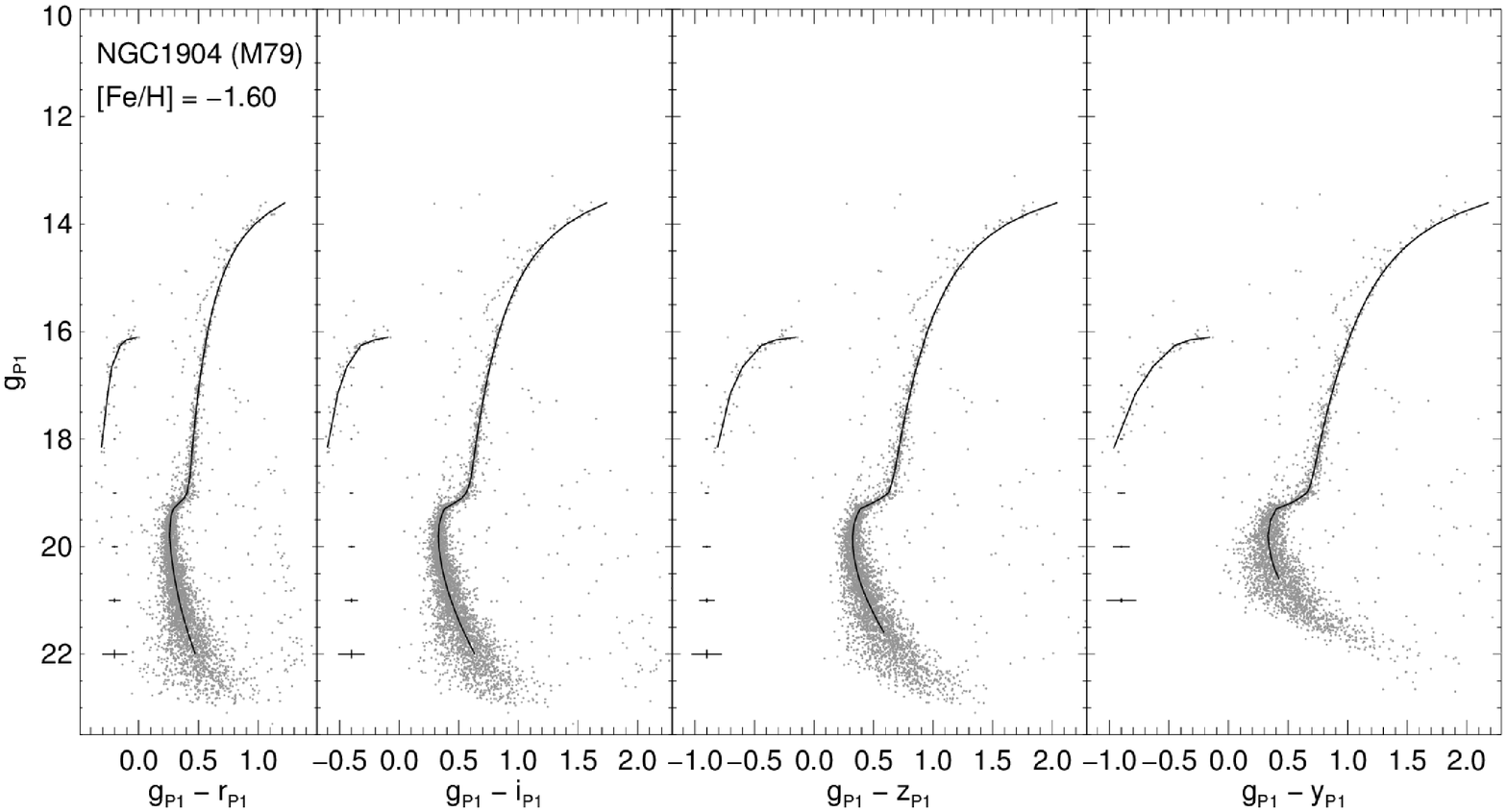}
\caption{Same as in Figure~\ref{fig:f288}, for NGC\,1904 (M\,79).}
\label{fig:f1904}
\end{figure*}

\begin{figure*}
\includegraphics[width=12.5cm]{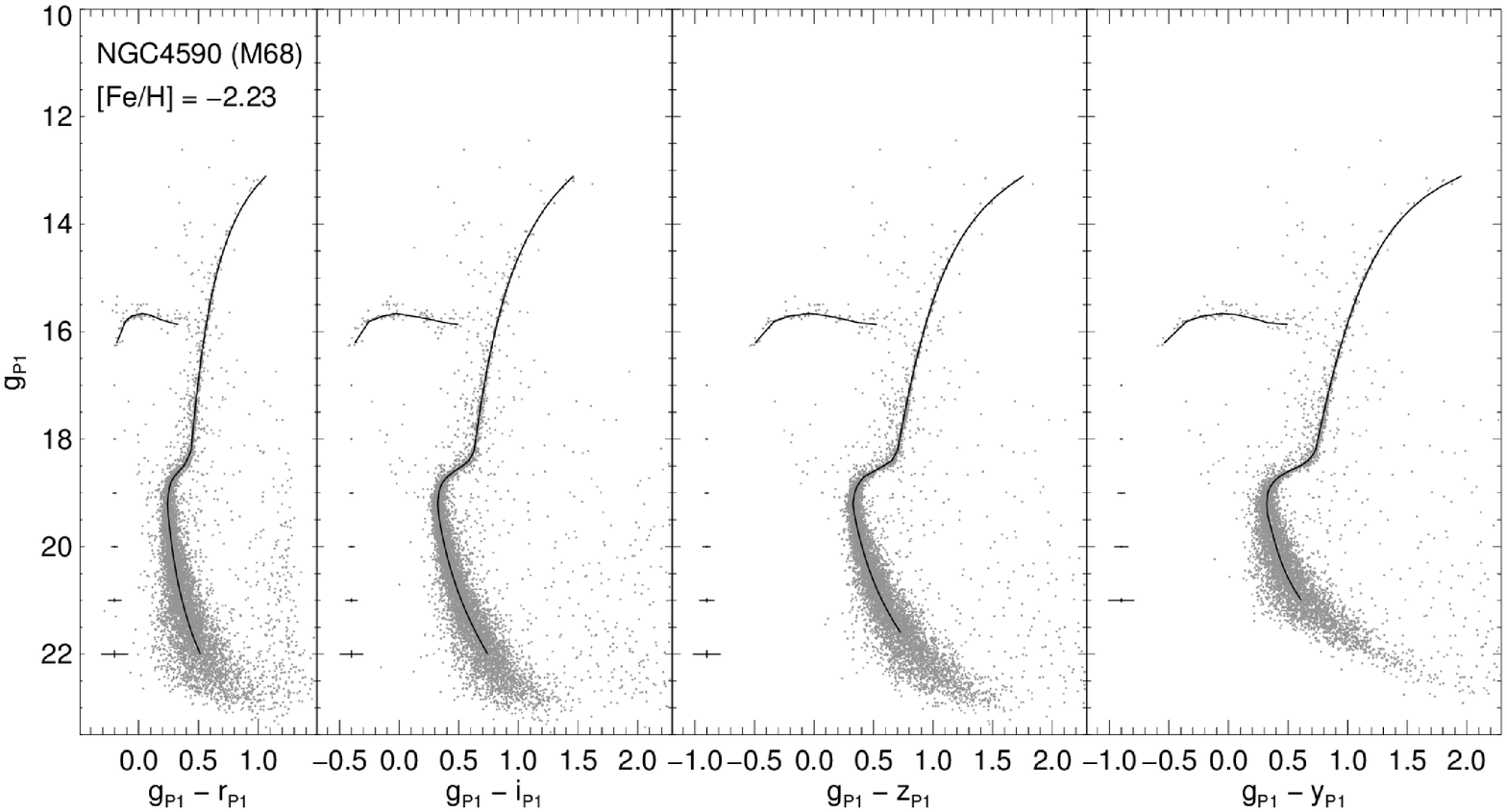}
\caption{Same as in Figure~\ref{fig:f288}, for NGC\,4590 (M\,68).}
\label{fig:f4590}
\end{figure*}

\begin{figure*}
\includegraphics[width=12.5cm]{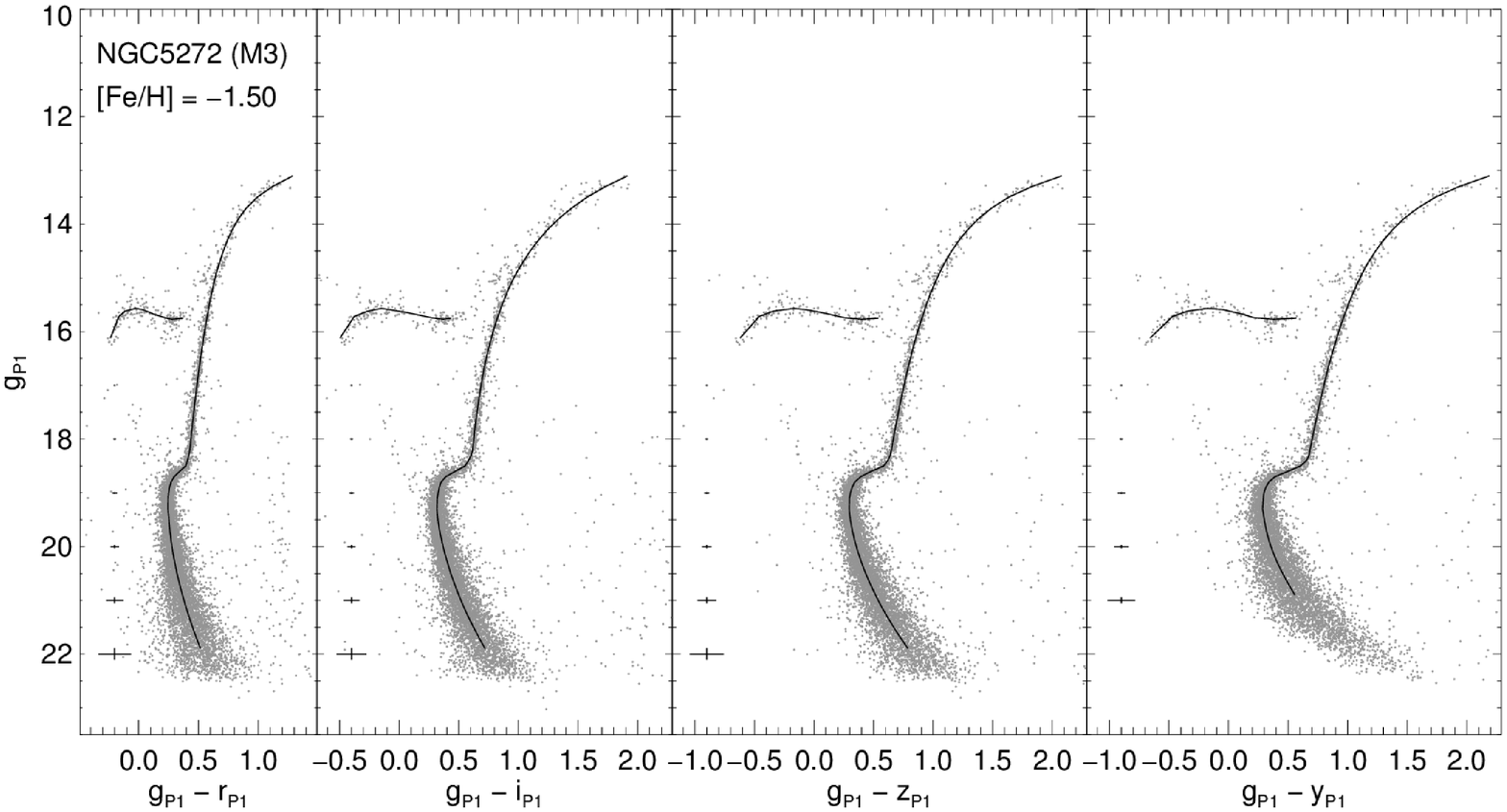}
\caption{Same as in Figure~\ref{fig:f288}, for NGC\,5272 (M\,3).}
\label{fig:f5272}
\end{figure*}

\begin{figure*}
\includegraphics[width=12.5cm]{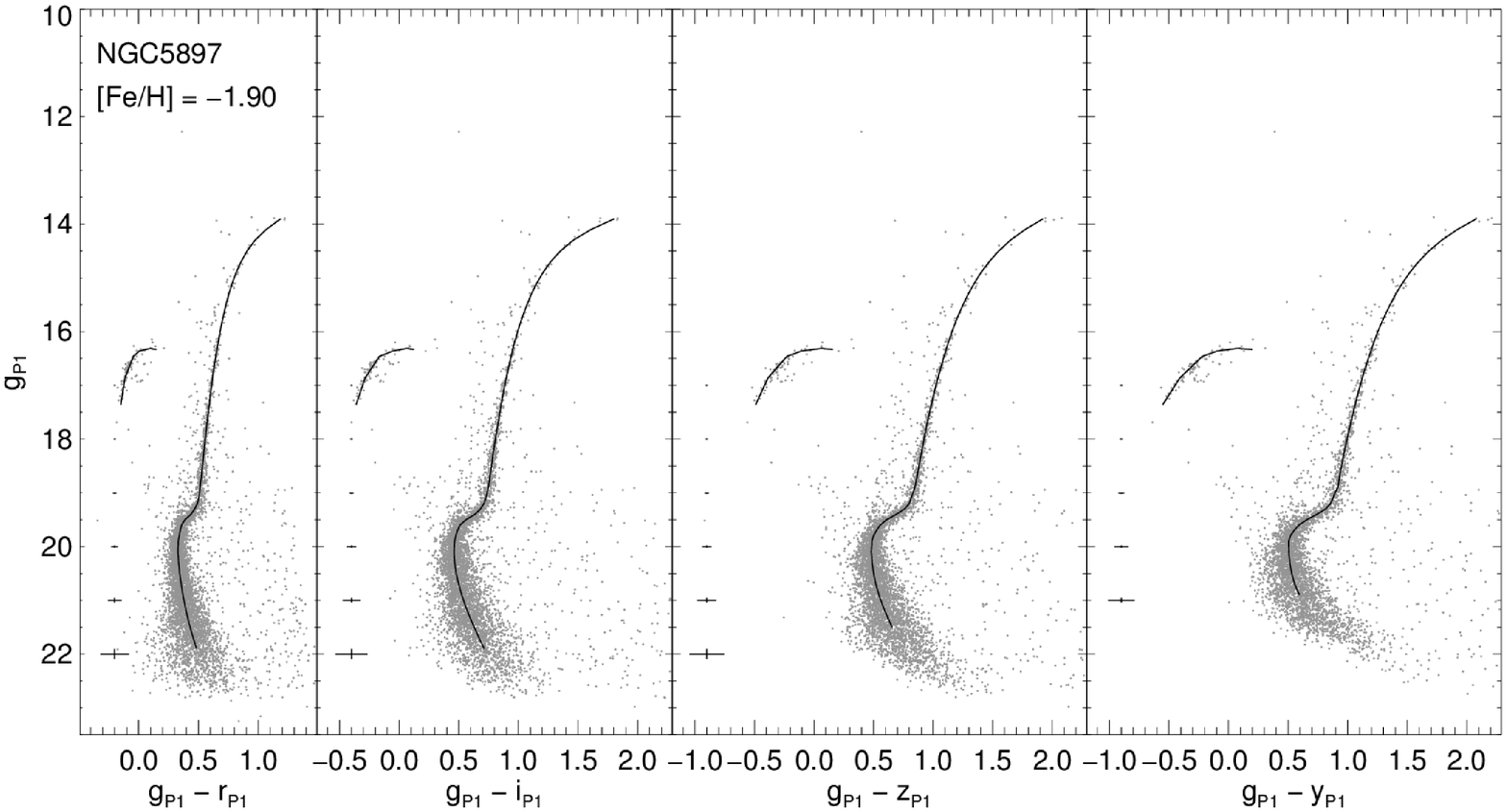}
\caption{Same as in Figure~\ref{fig:f288}, for NGC\,5897.}
\label{fig:f5897}
\end{figure*}

\begin{figure*}
\includegraphics[width=12.5cm]{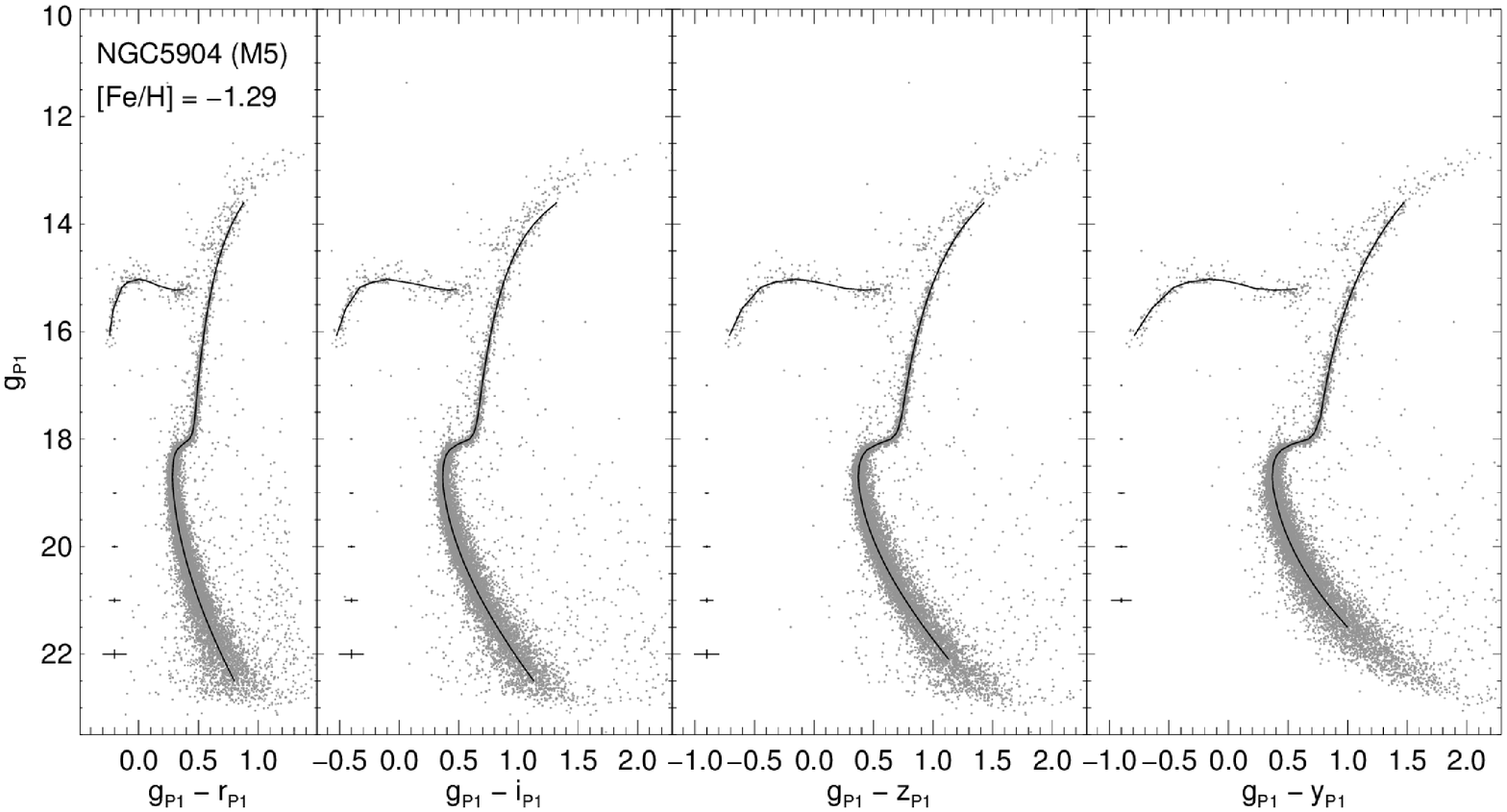}
\caption{Same as in Figure~\ref{fig:f288}, for NGC\,5904 (M\,5).}
\label{fig:f5904}
\end{figure*}

\begin{figure*}
\includegraphics[width=12.5cm]{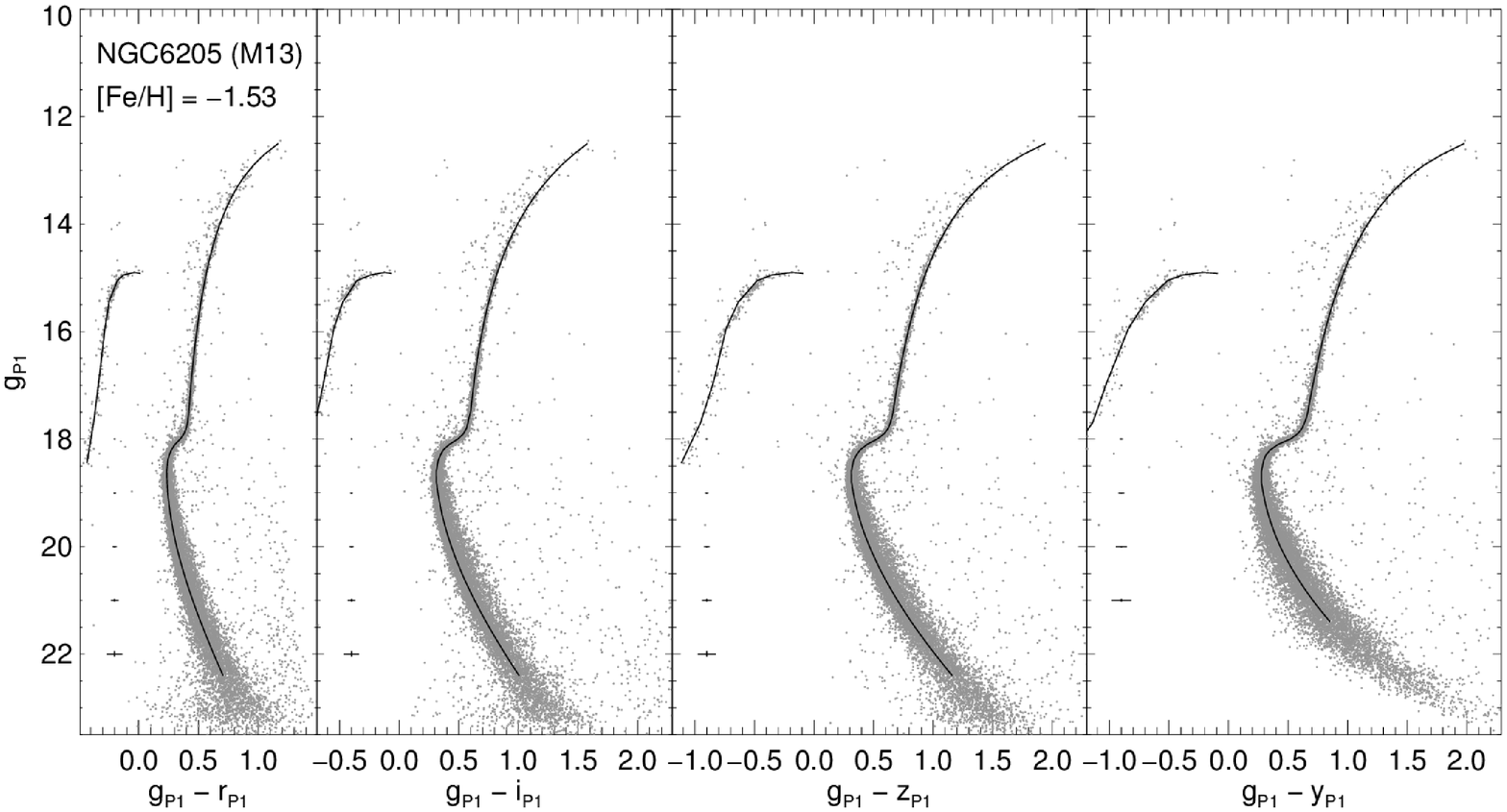}
\caption{Same as in Figure~\ref{fig:f288}, for NGC\,6205 (M\,13).}
\label{fig:f6205}
\end{figure*}

\begin{figure*}
\includegraphics[width=12.5cm]{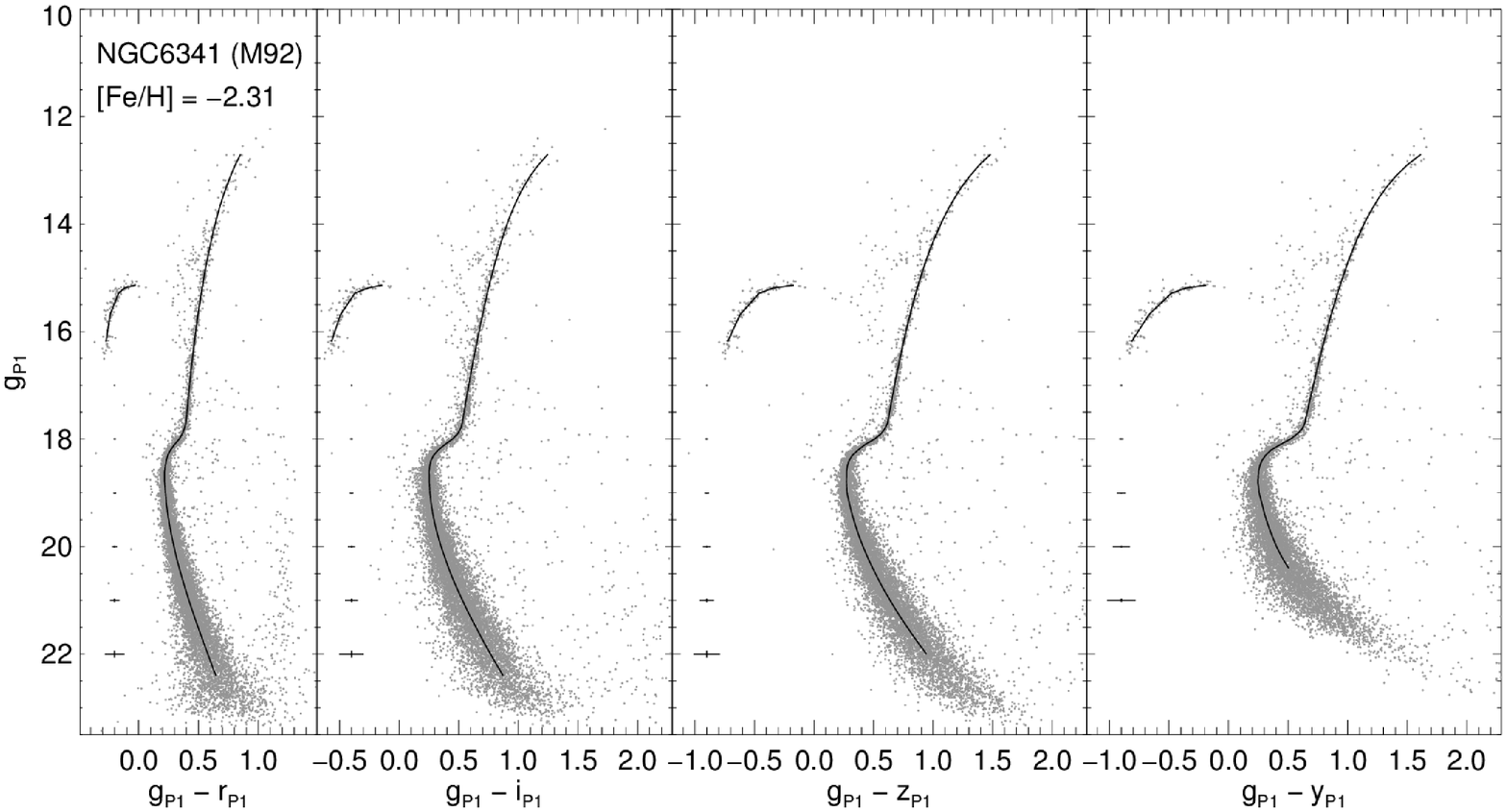}
\caption{Same as in Figure~\ref{fig:f288}, for NGC\,6341 (M\,92).}
\label{fig:f6341}
\end{figure*}

\begin{figure*}
\includegraphics[width=12.5cm]{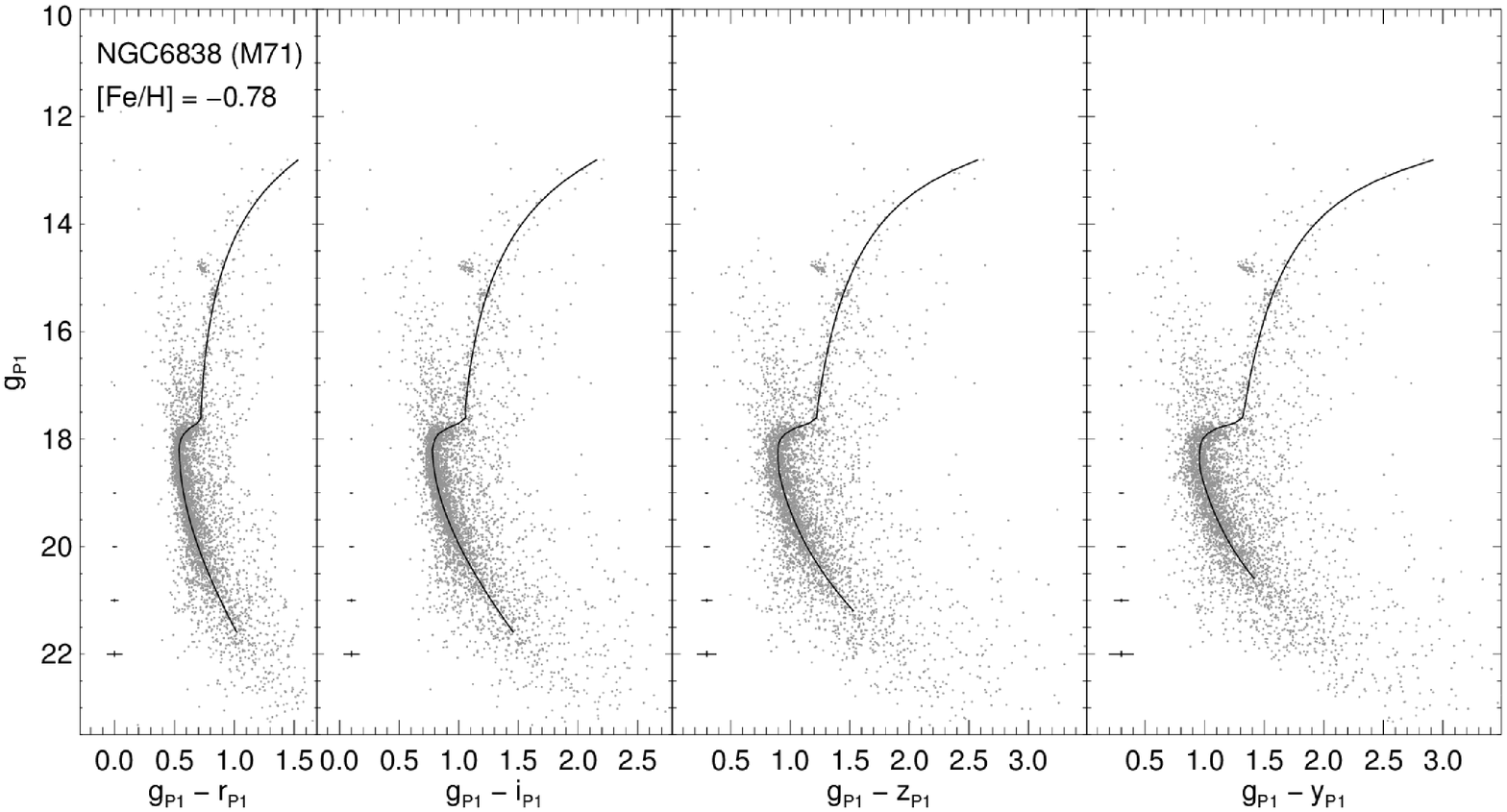}
\caption{Same as in Figure~\ref{fig:f288}, for NGC\,6838 (M\,71).}
\label{fig:f6838}
\end{figure*}

\begin{figure*}
\includegraphics[width=12.5cm]{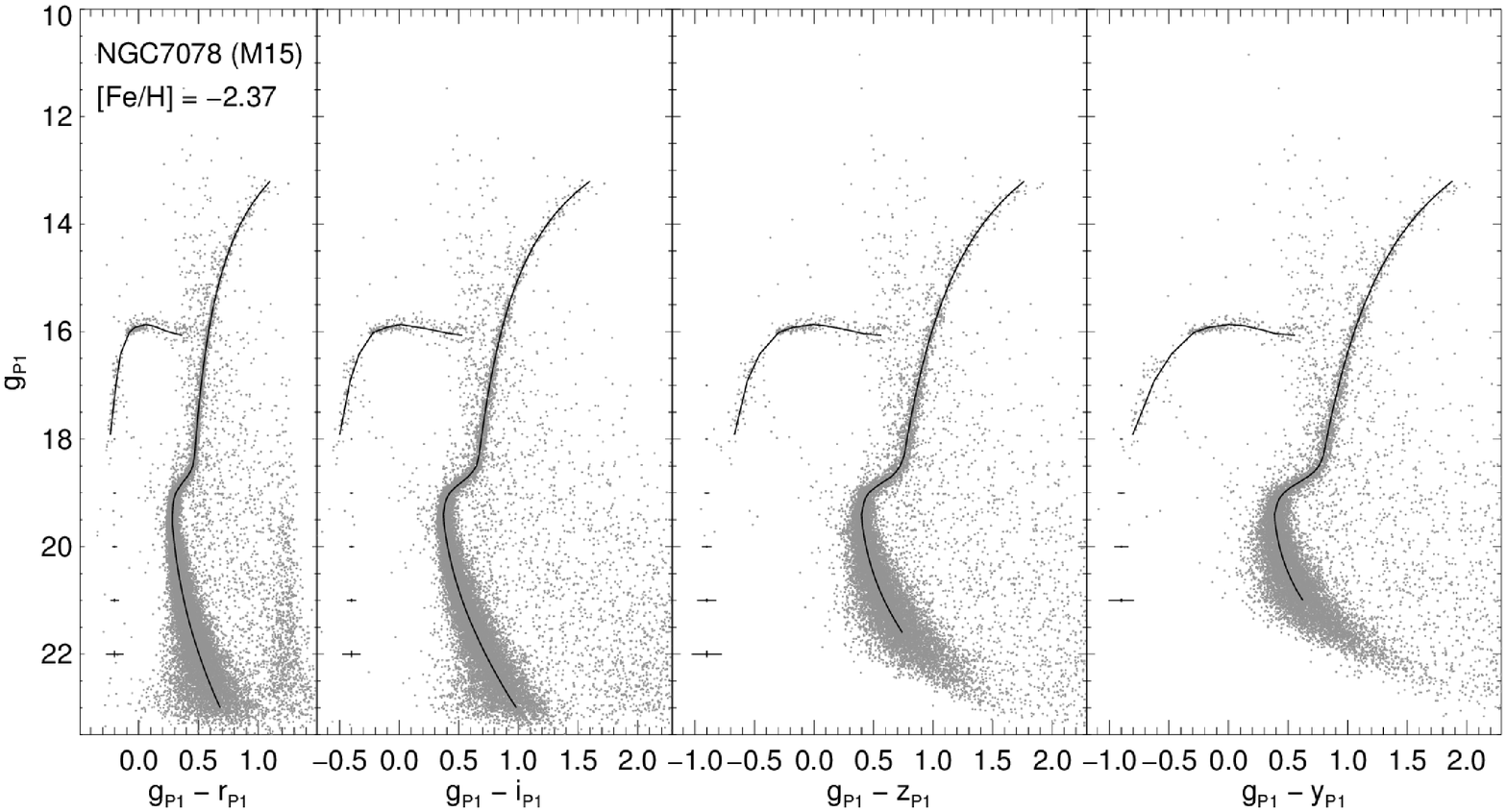}
\caption{Same as in Figure~\ref{fig:f288}, for NGC\,7078 (M\,15).}
\label{fig:f7078}
\end{figure*}

\begin{figure*}
\includegraphics[width=12.5cm]{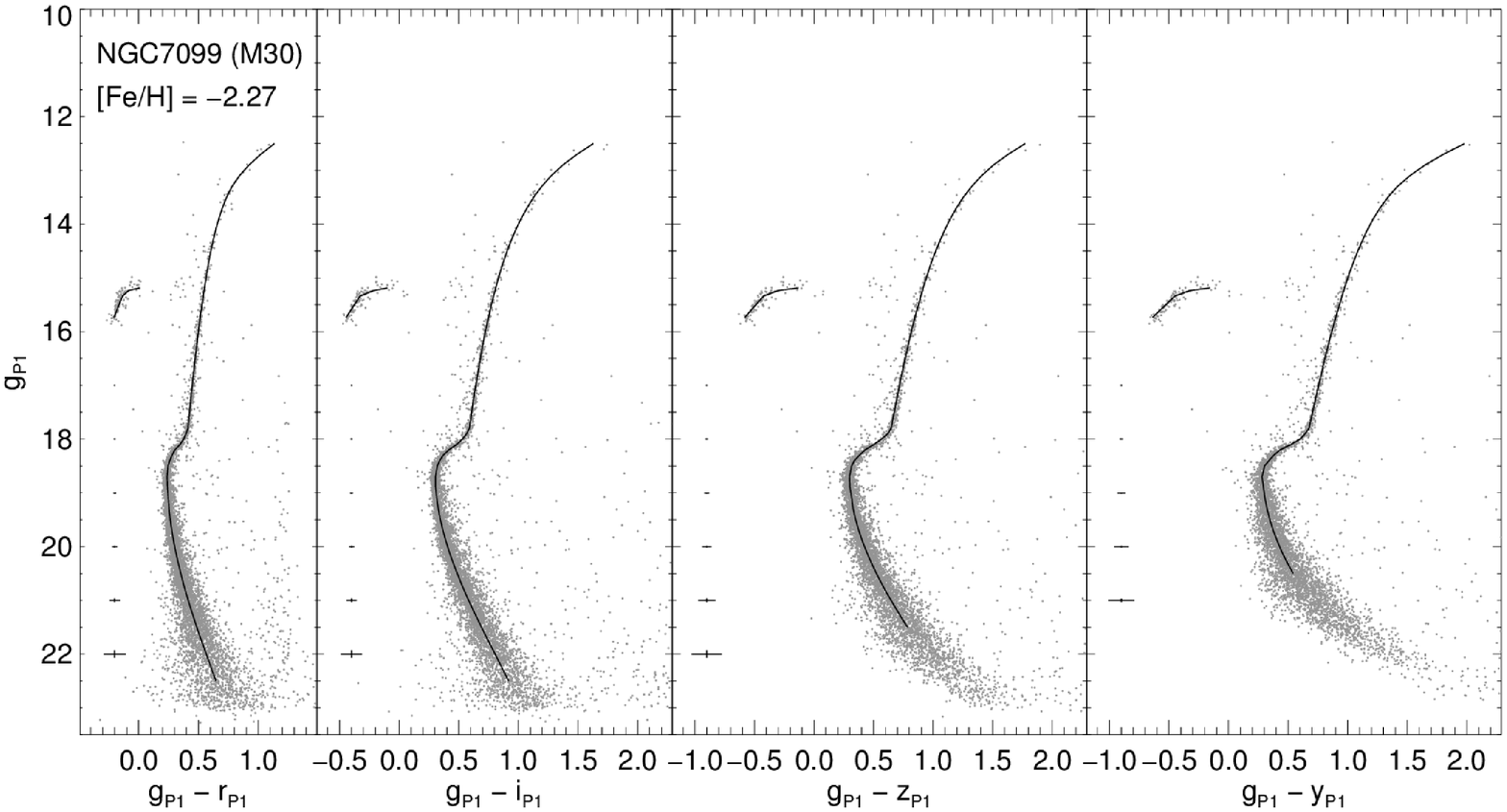}
\caption{Same as in Figure~\ref{fig:f288}, for NGC\,7099 (M\,30).}
\label{fig:f7089}
\end{figure*}

\begin{figure*}
\includegraphics[width=12.5cm]{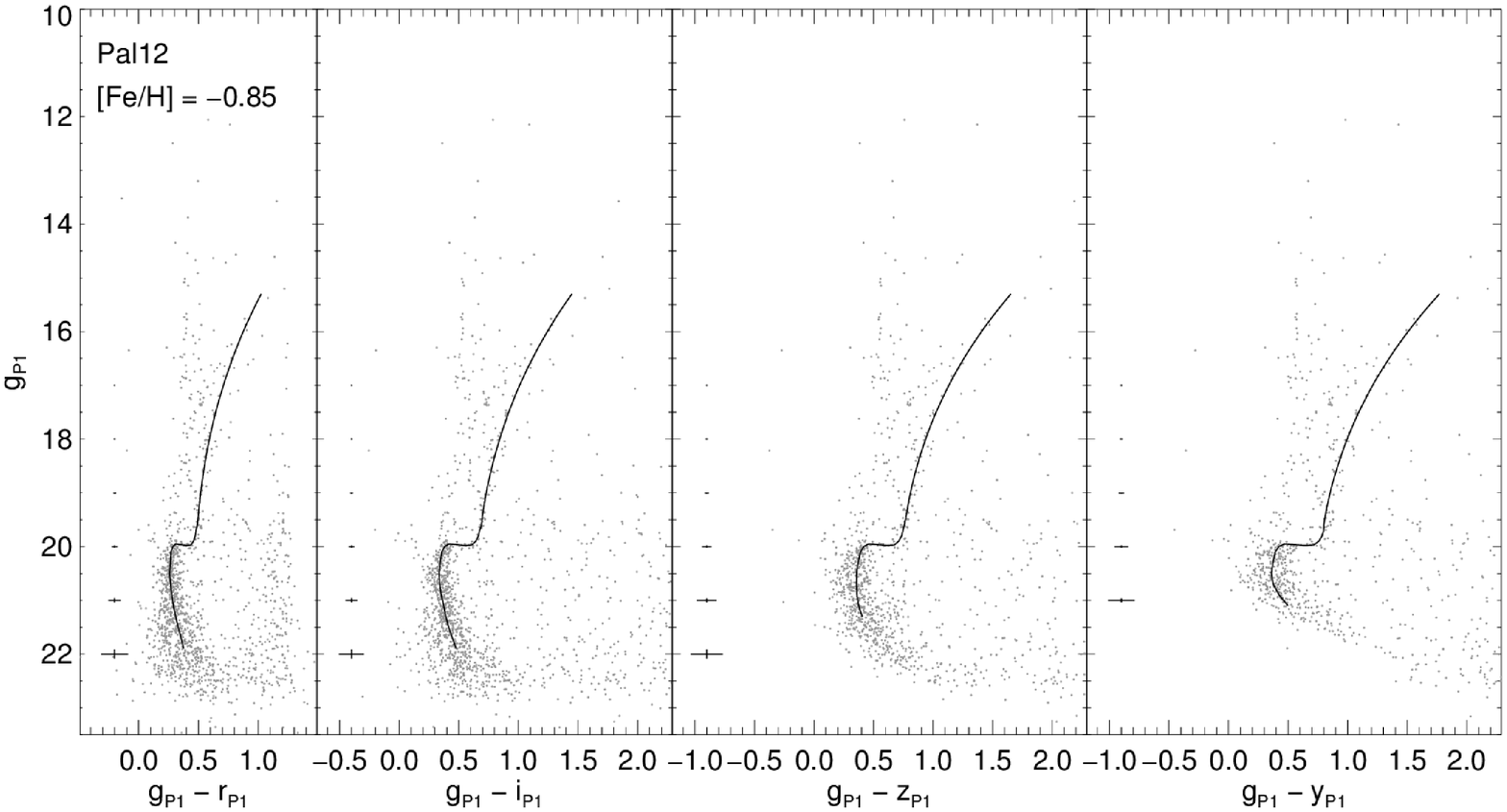}
\caption{Same as in Figure~\ref{fig:f288}, for Pal\,12.}
\label{fig:f12}
\end{figure*}

\begin{figure*}
\includegraphics[width=12.5cm]{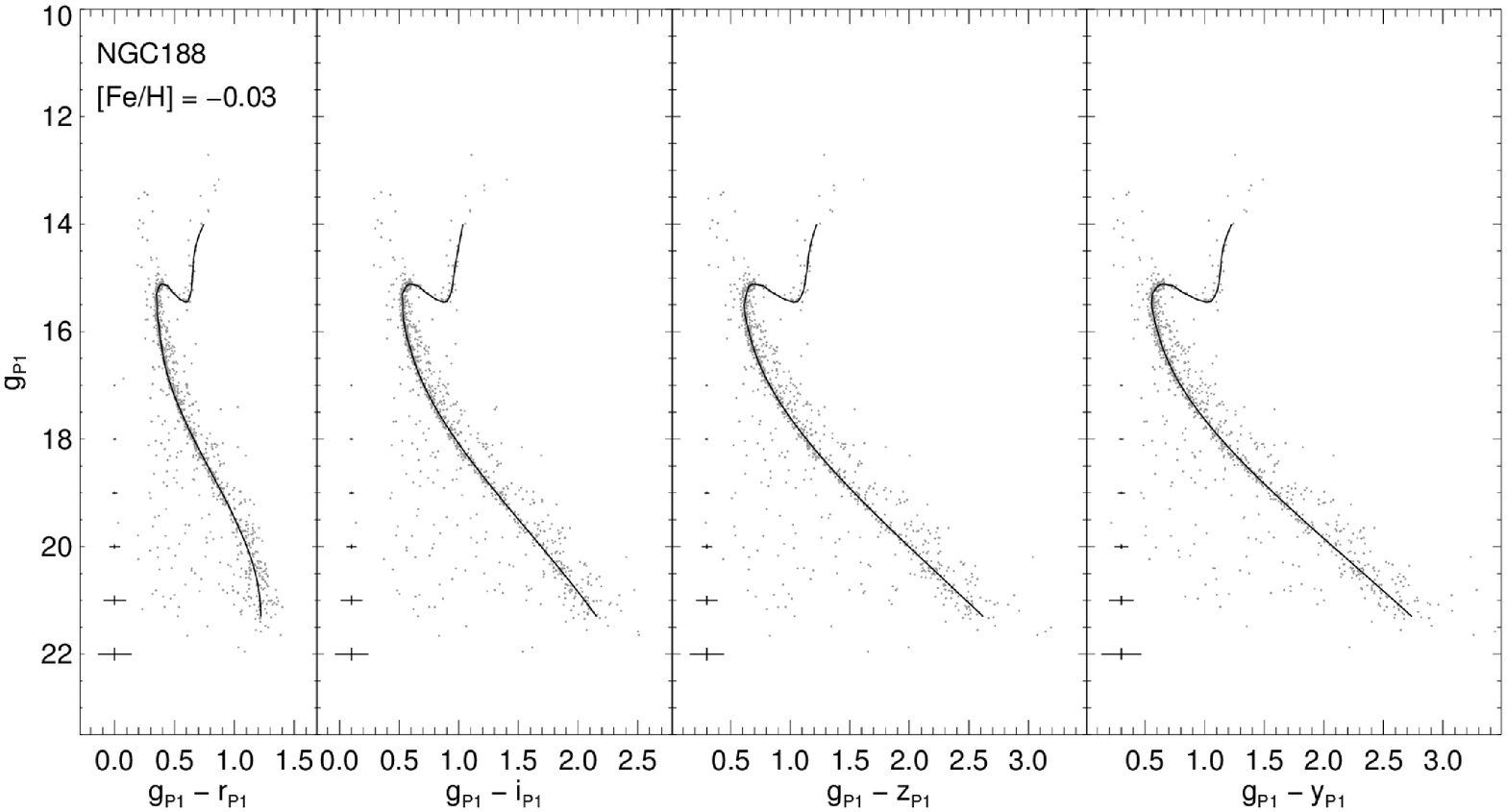}
\caption{Same as in Figure~\ref{fig:f288}, for open cluster NGC\,188.}
\label{fig:f188}
\end{figure*}

\begin{figure*}
\includegraphics[width=12.5cm]{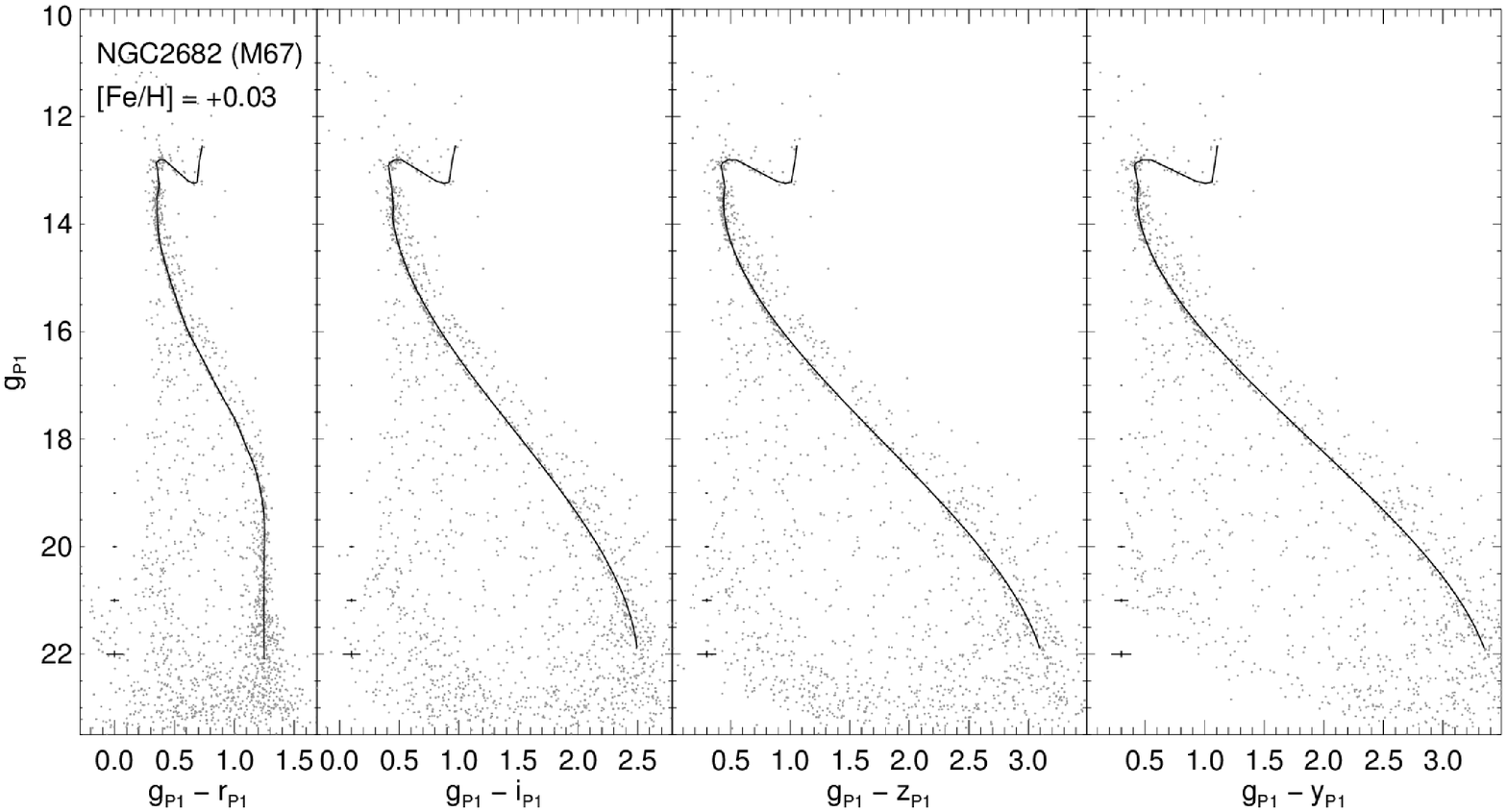}
\caption{Same as in Figure~\ref{fig:f188}, for NGC\,2682 (M\,67).}
\label{fig:f2682}
\end{figure*}

\begin{figure*}
\includegraphics[width=12.5cm]{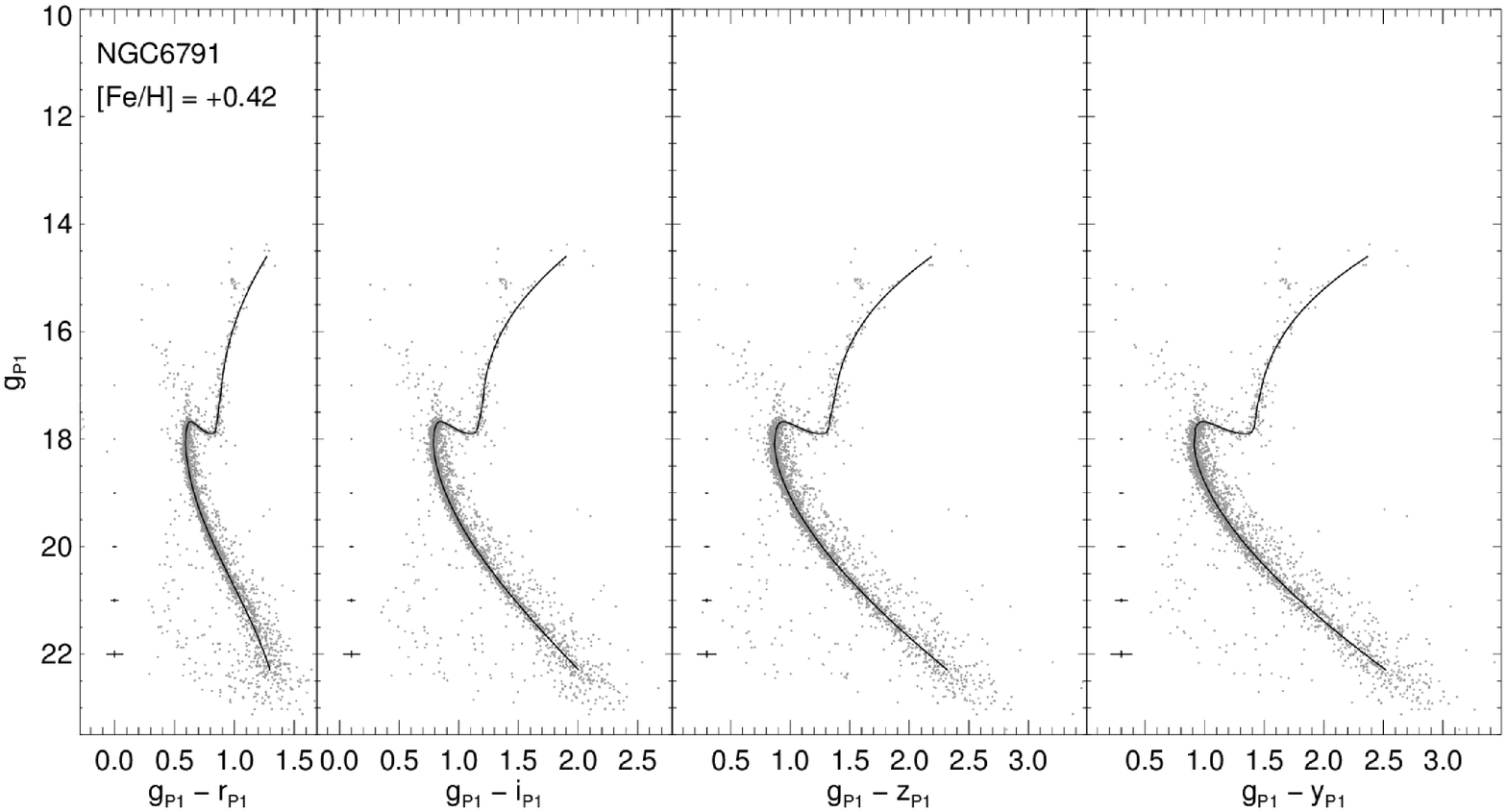}
\caption{Same as in Figure~\ref{fig:f188}, for NGC\,6791.}
\label{fig:f6791}
\label{lastpage}
\end{figure*}

 Figures \ref{fig:f288}--\ref{fig:f6791} present the CMDs and corresponding
 fiducials for the full sample of clusters analysed in this work. The
 fiducials are also provided in tabular form available as Supporting
 Information with the online version of the paper.

\end{document}